\documentclass[aps,prl,twocolumn,superscriptaddress,amsmath,amssymb]{revtex4-1}

\usepackage{graphicx}
\usepackage{multirow}
\usepackage{natbib}
\usepackage{array}
\usepackage{multirow,amssymb,amsbsy,amsmath,epstopdf}
\usepackage{color}

\usepackage[colorlinks,citecolor=blue]{hyperref}

\begin{document}
\title{Widening the sharpness modulation region of entanglement-assisted sequential quantum random access code: theory, experiment and application}

\author{Ya Xiao}\email{xiaoya@ouc.edu.cn}
\affiliation{Department of Physics, Ocean University of China, Qingdao 266100, People's Republic of China}
\author{Xin-Hong Han}
\affiliation{Department of Physics, Ocean University of China, Qingdao 266100, People's Republic of China}
\author{Xuan  Fan}
\affiliation{Department of Physics, Ocean University of China, Qingdao 266100, People's Republic of China}
\author{Hui-Chao Qu}
\affiliation{Department of Physics, Ocean University of China, Qingdao 266100, People's Republic of China}
\author{Yong-Jian Gu}\email{yjgu@ouc.edu.cn}
\affiliation{Department of Physics, Ocean University of China, Qingdao 266100, People's Republic of China}

\date{\today}

\begin{abstract}
Sequential quantum random access code (QRAC), allows two or more decoders to obtain desired message with higher success probability than the best classical bounds by appropriately modulating the measurement sharpness. Here, we propose an entanglement-assisted sequential QRAC  protocol which can enable device-independent tasks. By relaxing the equal sharpness and mutual unbiased measurement limits, we widen the sharpness modulation region from a one-dimensional interval to a two-dimensional triangle. Then, we demonstrate our scheme experimentally and get more than 27 standard deviations above the classical bound even when both decoders perform approximately projective measurements. We use the observed success probability to quantify the connection among sequential QRAC, measurement sharpness, measurement biasness and measurement incompatibility. Finally, we show that our protocol can be applied to sequential device-independent randomness expansion and our measurement strategy can enhance the success probability of decoding the entire input string. Our results may promote a deeper understanding of the relationship among quantum correlation, quantum measurement and quantum information processing.
\end{abstract}

\maketitle

\section*{I. Introduction}
The encoding-decoding process is of vital importance in security communication. How information can be encoded in a physical system and how much information can be retrieved lie at the core of communication theory. Random access code (RAC) is a communication protocol which allows us to encode a $m$-bit-long message into a shorter $n$ bits such that any one of the $ m $ bits can be recovered with a success probability no less than $1/2$. It has been shown that the probability can be increased if qubits are employed in encoding messages instead of classical bits \cite{ANT1999}. Such quantum random access code (QRAC) was first introduced for a simple preparation-and-measurement qubit system \cite{ANT1999, HIN2006,SBK2009,ALM2009}, and later developed for entanglement \cite{PZ2010, WWL2019} and higher-dimensional quantum system \cite{THM2015}. The diversity of QRAC protocol provides a wide range of applications in network coding \cite{HIN2007}, quantum key distribution \cite{PB2011,CVP2018}, random number generation \cite{LPY2012}, preparation contextuality \cite{SBK2009}, dimension witnessing \cite{AFM2018}, self-testing \cite{TKV2018,FK2019,AMC2020} and so on. 

To date, most discussions on decoding operator of QRAC protocol have been focusing on sharp (projective) measurement. This is because the sharper the measurement, the more information is obtained to decode a desired message. However, projective measurement may completely destroy the initial encoded state, so that at most one decoder can perform better than the best classical RAC. Recently, sequential unsharp (weak) measurement has attracted much attention. It has already been shown that by appropriately modulating the measurement sharpness, one can obtain enough information without disturbing the initial quantum correlation severely and others are capable of sharing the correlation simultaneously \cite{SGG2015,HZH2018}. Notably, it has been demonstrated theoretically and experimentally that multiple independent parties can share the nonlocality of a single maximally entangled qubit pair \cite{FCT2020,BC2020,RFY2020}. Moreover, unsharp measurements have been applied in quantum state tomography \cite{KKL2018}, incompatible observables \cite{PAL2016}, contextuality \cite{P2014,PALL2016}, and generate a long sequential sharing of other types of quantum correlations such as entanglement \cite{BMS2018}, steering \cite{CHP2020}, and quantum coherence \cite{DM2019}, etc.

These advances make it possible to develop protocols for multiple QRAC inequalities violation over sequential measurements, which was confirmed by Mohan, Tavakoli, and Brunner (MTB) \cite{MTB2019}. They found that two sequential decoders who can successfully decode message with a nontrivial probability beat the classical limits in $ 2\rightarrow 1 $ QRAC scenario. In MTB's protocol, the encoder Alice encodes her two classical bits into one qubit and sends it to the first decoder Bob. Then Bob applies quantum measurements unsharply, records the results and relays the resulting post-measurement states to the second decoder Charlie, who performs projective measurements on the received qubits. For Bob to successfully decode a message with a nontrivial probability beyond classical limits, the measurements must remain sharp; on the other hand, it should also be unsharp enough to retain some information for Charlie's measurement to make another QRAC inequality violation possible. In 2019, Mohan \textit{et al.} derived an optimal trade-off relation between these two QRACs, finding that both QRACs can outperform the optimal classical bound when Bob's measurement sharpness $\eta\in (1/\sqrt{2},(2+\sqrt{2})/4)$, which has been demonstrated experimentally \cite{AMC2020,FCV2020}.
 
It should be noted that there are some drawbacks in MTB's protocol which narrow the double QRAC inequality violation region. Firstly, in the choice of Bob's measurement direction, only the maximization of the probability between Alice and Bob ($ P_{AB} $) is considered, which will unavoidably lower the upper bound of the probability between Alice and Charlie ($ P_{AC} $). In fact, both QRAC probabilities should be considered to achieve a double QRAC inequality violation. Secondly, there is an implicit assumption that the sharpness of Bob's two measurements is equal. However, the adoption of unequal sharpness measurements may be more favorable; for instance, increasing the sequential length, i.e., more Bobs are allowed to share quantum correlation with Alice \cite{BC2020}. 

In this work, we firstly extend MTB's protocol into an entanglement-assisted scheme since entanglement-assisted QRAC performs better than a single photon QRAC  \cite{PZ2010,WWL2019,HSM2017} and has application for full device-independent quantum information task \cite{PAM2010}. Then, we introduce a method to widen the useful region of measurement sharpness parameter from a one-dimensional interval to a two-dimensional triangle to achieve double QRAC inequality violation. Experimentally, we verify the robustness and efficiency of our scheme and further illustrate how it works by quantifying the relationship among the degree of measurement sharpness, biasness and incompatibility with the observed two QRAC success probabilities. Finally, we present two specific applications of our protocol: sequential device-independent randomness expansion and increase the probability of successful decoding of the whole encoded string. 


\section*{II. Theoretical Model}


\begin{figure}[!h]
\centering\includegraphics[width=9cm]{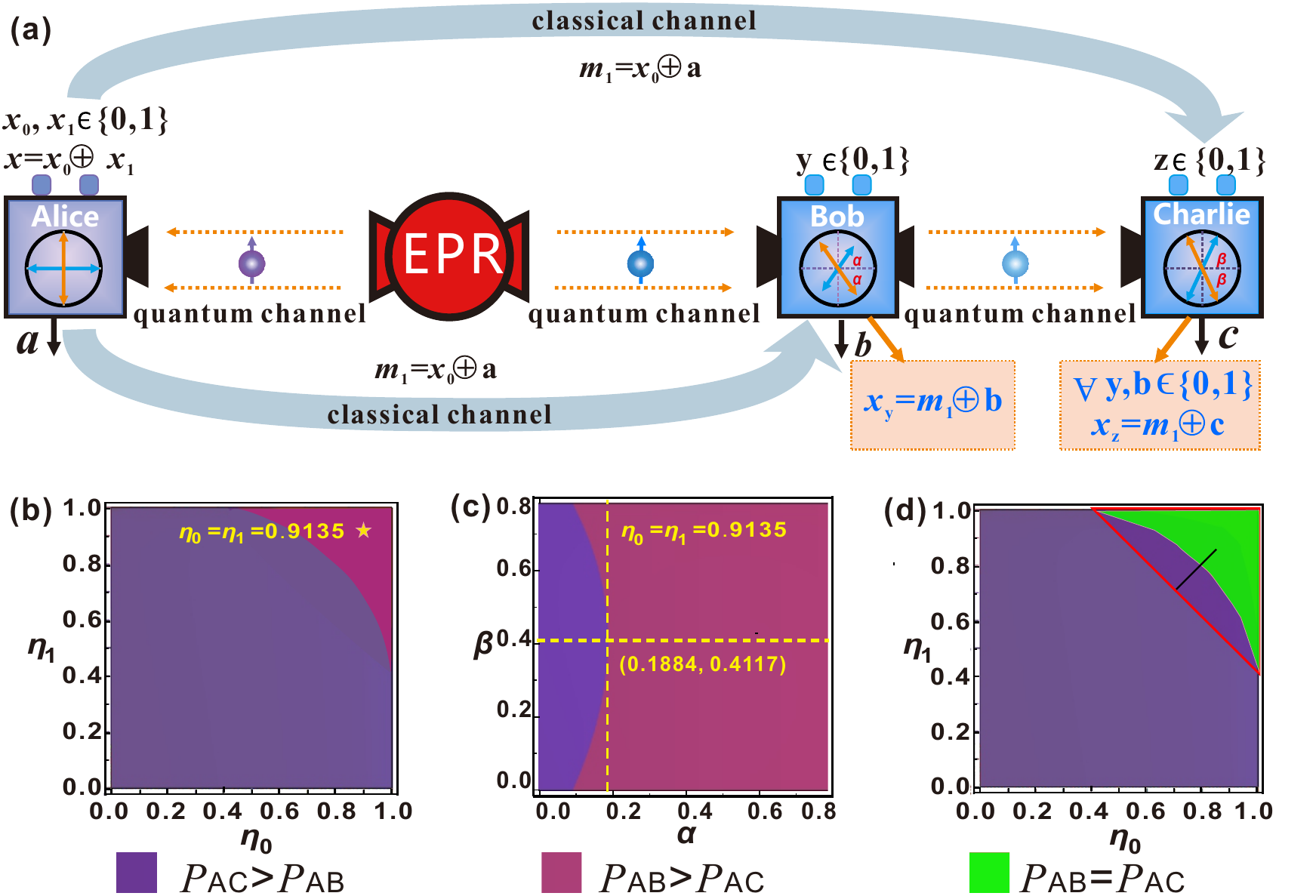}
\caption{(a). Entanglement-assisted quantum random access code via sequential measurement. The inserted circles represent our measurement settings, Bob and Charlie implement a separate QRAC with Alice. (b) the relationship between the average success probabilities $ \lbrace  P_{AB},P_{AC}\rbrace $ and Bob's two sharpness parameters $ \lbrace  \eta_{0}, \eta_{1}\rbrace $ when the measurement directions of Bob and Charlie both are set at $\alpha=\beta=\pi/4 $. (c). The optimal choice of measurements for Bob and Charlie when $\eta_{0}=\eta_{1}=0.9135$ (marked as star in (b)). (d) the relationship between the average success probabilities $ \lbrace  P_{AB},P_{AC}\rbrace $ and Bob's two sharpness parameters $ \lbrace \eta_{0}, \eta_{1},\rbrace $ when Bob and Charlie further optimizing their measurement directions by maximizing the minimum value of $\lbrace P_{AB},P_{AC} \rbrace $. The area enclosed by the red curve correspond to the double QRAC inequality violation area, i.e., $ P_{AB}\geqslant 3/4$ and $P_{AC}\geqslant 3/4 $, in our measurement strategy. The black line represents the range of MTB's measurement strategy. See Fig. \ref{range}(a) for more details.}
\label{sqmodel}
\end{figure}
We start with a brief review of MTB's QRAC protocol. In their protocol, Alice encodes her two classical bits into one of four pure states, which form the vertices of a square in the equatorial line of the Bloch sphere, and then sequentially transmits it to Bob and Charlie to measure \cite{MTB2019}. Here, we extent it into an entanglement-assisted scheme where the encoding messages are hidden in the entanglement state shared among Alice and Bob (Charlie) rather than in a proper qubit. The process of communication between Alice and Bob (Charlie) is completed by measuring each qubit on their own part \cite{WWL2019}.

To clarify our entanglement-assisted sequential QRAC scheme, we show the encoding-decoding process in the case of $ 2\rightarrow 1$ in Fig. \ref{sqmodel}(a). Two decoders, Bob and Charlie, attempt to recover a desired bit encoded in Alice's string with a non-trivial success probability using only a pair of entangled particles. We denote the measurement direction of Alice, Bob, and Charlie by the Bloch vector $\vec{r}_{x}$, $\vec{s}_{y}$ and $\vec{t}_{z}$, respectively. To begin with, a pair of two-qubit entanglement particles $\rho_{AB}$ are shared between Alice and Bob. Alice encodes her two classical bits, $ x_{0}, x_{1}\in \lbrace 0,1 \rbrace$, by performing a local measurement $  M_{a\vert\vec{r}_{x}} $ on her particle based on the value of $ x= x_{0}\oplus x_{1}$. The corresponding measurement result is denoted as $a \in \lbrace 0,1 \rbrace$. After the measurement, she separately sends one classical bit $ m_{1}= x_{0}\oplus a $ to Bob and Charlie, which will help them obtain a higher success decoding probability. In the decoding process, Bob chooses an unsharp measurement $ M_{b\vert\vec{s}_{y}}=K^{\dagger}_{b\vert\vec{s}_{y}}K_{b\vert\vec{s}_{y}} $ according to an input $y \in \lbrace 0,1 \rbrace$ to decode a desired bit $ x_{y}$. $ K_{b\vert\vec{s}_{y}}$ is the corresponding Kraus operator. Then, he records the output result $b \in \lbrace 0,1 \rbrace$ and compares it with the received bit $ m_{1} $. The decoding process is successful if $ x_{y}= m_{1}\oplus b $, and the success probability can be calculated by $P(x_{y}= m_{1}\oplus b\vert x,y)=Tr[ M_{a\vert\vec{r}_{x}}\otimes M_{b\vert\vec{s}_{y}}\cdot\rho_{AB}]$. Similarly, Charlie receives an input $z \in \lbrace 0,1 \rbrace$ with an aim to guess $ x_{z}$  by taking an optimal measurement $ M_{c\vert\vec{t}_{z}} $ on the particle he received from Bob, yielding an outcome $c \in \lbrace 0,1 \rbrace$. Since Bob and Charlie separately implement a QRAC with Alice, they act independently. It means Charlie is unaware about what measurements are performed by Bob, thus we have to take the average of Bob's input and output to obtain the state shared between Alice and Charlie, which is given by $\rho_{AC}= \dfrac{1}{2}\sum \limits_{y,b}( I_{A}\otimes K_{b\vert\vec{s}_{y}})\rho_{AB}( I_{A}\otimes K^{\dagger}_{b\vert\vec{s}_{y}})$.  Charlie's success probability  then can be given as $P(x_{z}= m_{1}\oplus c\vert x,z)=Tr[ M_{a\vert\vec{r}_{x}}\otimes M_{c\vert\vec{t}_{z}}\cdot\rho_{AC}]$. Suppose all the inputs $(x_{0},x_{1},y,z)$ are statistically independent and uniformly distributed. Repeat the process many times, the average success probability of Bob and Charlie, $P_{AB} $ and $ P_{AC}$, can be obtained by 
\begin{equation}\label{pro1}
\begin{split}
&P_{AB}=\dfrac{1}{8}\sum \limits_{x,y}P(x_{y}= m_{1}\oplus b\vert x,y), \\
&P_{AC}=\dfrac{1}{8}\sum \limits_{x,z}P(x_{z}= m_{1}\oplus c\vert x,z).\\
\end{split}
\end{equation}
The optimal classical limit of both average success probabilities is $3/4$ \cite{ANT1999}. When Alice and Bob (Charlie) behave classically, then $P_{AB}\leq 3/4$ ($P_{AC}\leq 3/4$), which we call QRAC inequality. It has been demonstrated that the double QRAC inequality violation can be observed simultaneously by adjusting Bob's measurement sharpness parameter and its useful range is $ (1/\sqrt{2},(2+\sqrt{2})/4)$ \cite{MTB2019}. In the following, we will show that the range can be expanded by optimizing the measurement settings of both decoders.

As the MTB's QRAC protocol, we assume Alice encodes her two classical bits in the $XOZ$ plane. For convenience, we suppose Alice performs measurement along $ X $ and  $ Z $ axes. It obviously that the optimal decoding operators need to be performed on the same plane \cite{TKV2018}. Thus, the decoding measurement direction for Bob and Charlie can be expressed as $cos \alpha X \pm sin\alpha Z$  and $cos\beta X \pm sin\beta Z$ respectively (as shown in Fig. \ref{sqmodel}(a)). Here, we relax the equal sharpness and mutual unbiased conditions of each decoder's respective pair of measurements in the previous schemes \cite{MTB2019,FCV2020,AMC2020}. Then the measurement settings of each encoder and decoder can be rewritten as

\begin{equation}\label{measurement}
\begin{split}
&M_{a\vert\vec{r}_{x}}=[I_{A}+(-1)^a ((x\oplus 1) \sigma_{1}+x\sigma_{3})]/2,\\
&M_{b\vert\vec{s}_{y}}=[I_{B}+(-1)^b \eta_{y} (cos\alpha\sigma_{1}+ (-1)^y sin\alpha \sigma_{3})]/2,\\
&M_{c\vert\vec{t}_{z}}=[I_{C}+(-1)^c (cos\beta\sigma_{1}+ (-1)^z sin\beta\sigma_{3})]/2,\\
\end{split}
\end{equation}
where $ \eta_{y}\in(0,1)$ is the sharpness parameter. With these measurement settings and an initial maximum entanglement state, the expected average success probability of Bob and Charlie, $P_{AB} $ and $ P_{AC}$, can be expressed as 
\begin{equation}\label{optiprob}
\begin{split}
&P_{AB} =\dfrac{1}{8}[4+(\eta_{0}+\eta_{1})\cdot(cos\alpha+sin\alpha)],\\
&P_{AC} =\dfrac{1}{8}[4+2\cdot M+N\cdot(\sqrt{1-\eta^{2}_{0}}+\sqrt{1-\eta^{2}_{1}})],\\
\end{split}
\end{equation}
where $ M=cos^{2}\alpha \cdot cos\beta+sin^{2}\alpha \cdot sin\beta $, $ N=cos^{2}\alpha \cdot sin\beta+sin^{2}\alpha \cdot cos\beta $, $ \alpha, \beta\in [0,\pi/4] $. Specifically, when $ \eta_{0}= \eta_{1}=\eta $ and $\alpha=\beta=\pi/4 $, our scheme goes back to an entanglement-assisted MTB's protocol with the same sharpness modulation range, $\eta\in (1/\sqrt{2},(2+\sqrt{2})/4)$.

It easy to find that, from from Eq. (\ref{optiprob}), for a given pair of values of $ \eta_{0}$ and $\eta_{1} $, once $ \alpha $ set, not only $ P_{AB} $ can be obtained, but also a proper value of $ \beta $ can be found to maximize the value of $ P_{AC} $. In order to keep both probabilities surpassing the optimal classical limit within a wider sharpness region, the best way is to choose a pair of $\lbrace \alpha,\beta\rbrace$ that maximizes the minimum value of $\lbrace P_{AB}, P_{AC}\rbrace$. Noting that the maximum value of $P_{AB}$ is always obtained at $\alpha=\pi/4 $ once $\eta_{0}$ and $\eta_{1}$ are set, and if $ \beta $ is updated to maximize $ P_{AC} $, $ P_{AC} $ will increase with a gradual decrease of $ \alpha $ from $ \pi/4 $ to $ 0 $, and $ P_{AB} $ will decrease at the same time. Thus, for each pair of $\lbrace \eta_{0},\eta_{1} \rbrace $, we start with $\alpha=\pi/4 $, then maximize the value of $ P_{AC} $, finally compare the values of $ P_{AB} $ and $ P_{AC} $. If $ P_{AB} \leq P_{AC}$, the current values of $\lbrace \alpha,\beta\rbrace$ are set as the optimal measurement parameters for Bob and Charlie. This is because that in this case, maximizes the minimum value of $\lbrace P_{AB}, P_{AC}\rbrace$ equals to maximize the value of $P_{AB}$ which can be realized by setting $\alpha=\pi/4 $, and in the meantime the maximum value of $P_{AC}$ can be obtain with $\beta=\pi/4 $. Clearly, when $ P_{AB} \leq P_{AC}$, both Bob's and Charlie's measurement directions are same as MTB's protocol; otherwise, if $ P_{AB} > P_{AC}$ when $ \alpha=\pi/4 $, we have to optimize their measurement directions to increase the value of $ P_{AC}$. Here, we gradually decrease the value of $ \alpha $, then choose a proper value of $ \beta $ to maximize $ P_{AC} $ for each given  $\alpha $. The maximum value of the minimum value of $\lbrace P_{AB}, P_{AC} \rbrace $ can be obtained when $ P_{AB}=P_{AC} $. Fig. \ref{sqmodel}(c) shows the optimal value of $ \alpha $ and  $ \beta $ in the case of $\eta_{0}=\eta_{1}=0.9135$. Similarly, for other given pairs of $ \lbrace \eta_{0},\eta_{1} \rbrace$, we can always find the optimal numerical solutions of $\lbrace \alpha, \beta \rbrace$, see Appendix A for more details. Eventually, we can further extend double QRAC inequality violation region as the area enclosed by the red curve line in Fig. \ref{sqmodel}(d), which is much larger than the original range $\eta_{0}=\eta_{1}\in (1/\sqrt{2},(2+\sqrt{2})/4)$, the black line. 


\section*{III. Experimental method}


\begin{figure}[!h]
\centering\includegraphics[width=9cm]{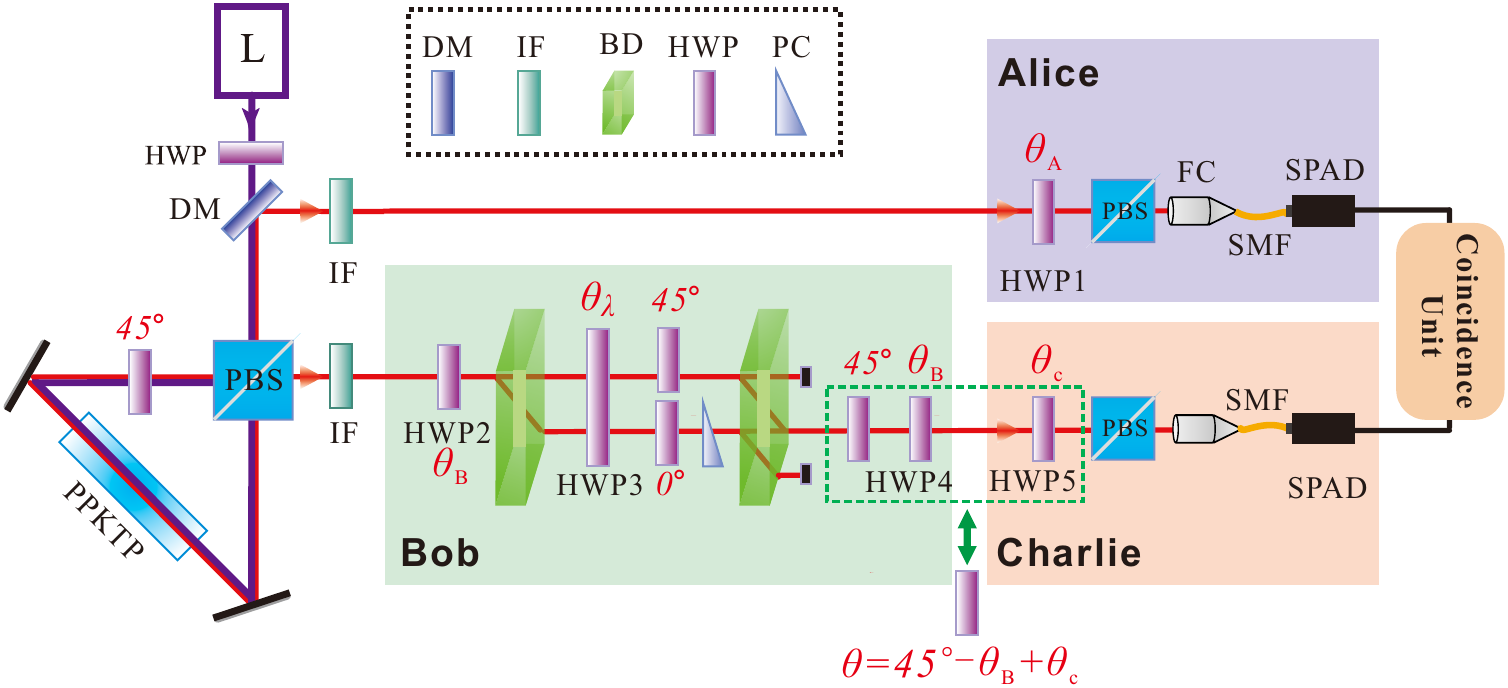}
\caption{Experimental setup. A bright source of polarization-entangled photon pairs in a state $(\vert HH \rangle + \vert VV \rangle )/\sqrt{2} $ is generated via the spontaneous parametric down-conversion process by pumping a type-II cut PPKTP crystal located in a Sagnac interferometer with an ultraviolet laser (L) at 405 nm. These two photons are filtered by interference filters  (IFs). One of the photons is sent to Alice, and the other one is subsequently sent to Bob and Charlie. A balanced Mach-Zehnder interferometer (MZI) consists of two calcite beam displacers (BDs) and a couple of half-wave plates  (HWPs) is used to realize Bob's unsharp measurement, where the sharpness parameter $\eta_{y}=cos(4 \theta_{\lambda}) $ can be tuned conveniently by rotating the internal HWP3. HWP2 and HWP4 are used to change any of $\lbrace M_{0\vert\vec{s}_{0}},M_{1\vert\vec{s}_{0}}, M_{0\vert\vec{s}_{1}}, M_{1\vert\vec{s}_{1}} \rbrace $ into another as well as selecting the basis and outcome of the measurement. The relative phase between the exist two paths can be controlled conveniently by employing removable wedge phase plate (PC). Both Alice and Charlie only perform projective measurements in the space of linear polarization, hence their setup can be simplified to a compose of a HWP and a polarization beam splitter (PBS). To reduce the number of components, we replace the group of three consecutive HWPs in the green dot box with a single HWP. The photons are detected by single-photon detectors (SPADs) and the signals are sent for coincidence.}
\label{setup}
\end{figure}


\begin{figure*}[!htp]
\centering\includegraphics[width=18cm]{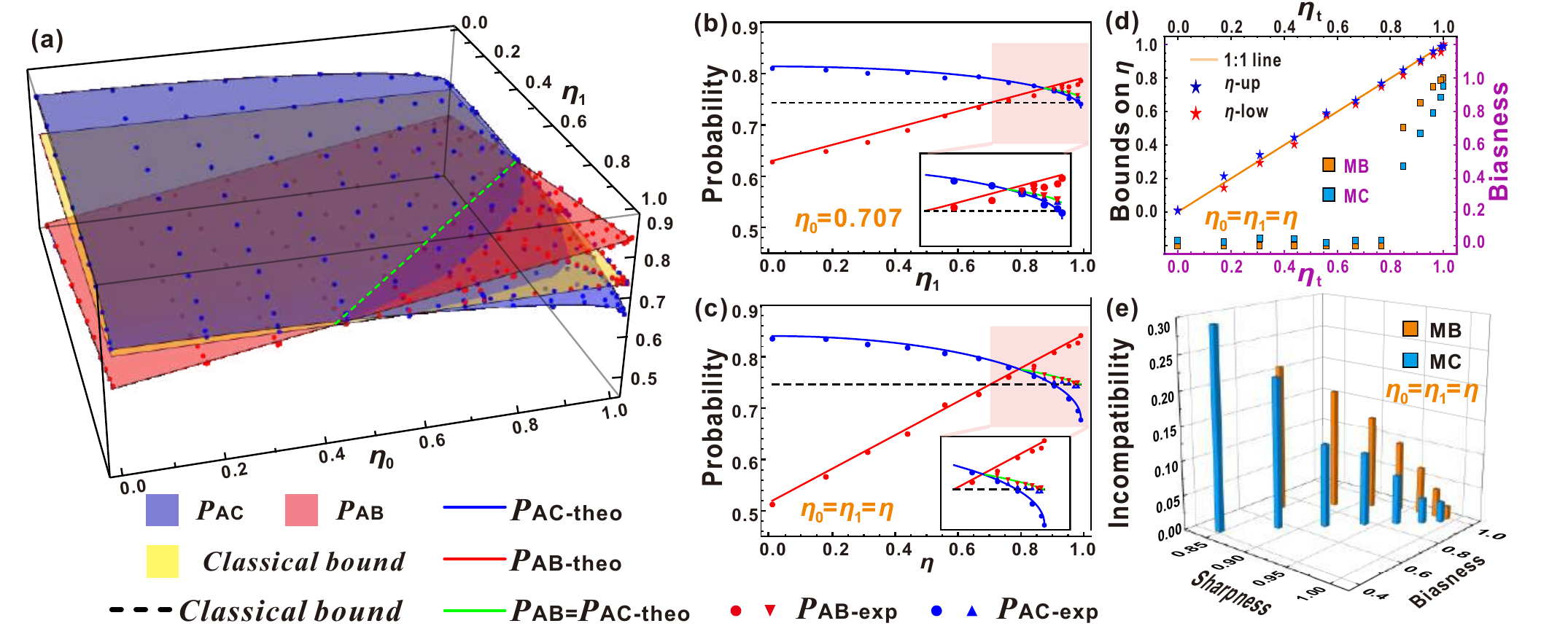}
\caption{Demonstrating the double violation of QRAC inequality. (a). The average success probabilities are experimentally measured for various pairs of measurement
sharpness parameters $ \lbrace \eta_{0} , \eta_{1} \rbrace$. Experimentally measured average success probabilities are shown as markers, where the red and blue circles (triangles) correspond to $ P_{AB} $ and  $ P_{AC} $ when Bob and Charlie have (have not) further optimizing their measurement directions at $P_{AB}>P_{AC}$. Theoretical predictions are also represented as surface plots, and compared with the experimental data. The average success probabilities plot as a function of $\eta_{1} $ for fixed $\eta_{0}=0.707 $ (b) and  $\eta_{0}=\eta_{1}=\eta $ (c). The shaded areas with light red color indicate the double violation of QRAC inequality. It is clearly demonstrated that the double violation area can be further expanded with the mutually biased measurements. (d). The lower and upper bounds of Bob's sharpness parameter (stars) together with the upper bound of both decoders measurement biasness (squares) obtained by the experimental average success probabilities vs the target measurement sharpness. It is obvious that with the increasing of the measurement sharpness, both decoders' biasness parameter increases but their difference decreases. (e). The relationship between the lower bound on the degree of incompatibility in Bob's (dark yellow) and Charlie's (dark cyan) mutually biased measurements corresponding to seven different values of sharpness and biasness parameters show in Fig. \ref{range}(c). They are both above zero in the optimizing measurement range and decrease as the sharpness and biasness parameter increase. Error bars are estimated by the Poissonian statistics of two-photon coincidences which are too small to be visible (see Appendix C for more details).}
\label{range}
\end{figure*}
 
To verify the advantage of our new protocol, we put it into practice with an optic experiment. Fig. \ref{setup} shows our experimental setup. A 405 nm continuous wave diode laser (L) is used to pump a 15 mm long PPKTP crystal inside a polarization Sagnac interferometer in clockwise and counter-clockwise direction to generate a maximum two-qubit polarization-entangled photons state $(\vert HH \rangle + \vert VV \rangle )/\sqrt{2}$ \cite{FHP2007}. Two interference filters(IFs) with bandwidth of 3 nm are used to filter these photons. One of the two photons is sent to Alice to encode her two classical bits, the other photon is sequentially sent to two decoders, Bob and Charlie.  During the encoding-decoding process, Alice encodes her two classical bits into her measurement by using a combination of halfwave plate (HWP1) and polarized beam splitter (PBS), shown in the light blue green region. The angle of HWP1 is set at  $22.5^\circ $, $-22.5^\circ $, $0^{\circ} $ and $ 45^\circ $ for $ M_{0\vert\vec{r}_{0}}$, $ M_{1\vert\vec{r}_{0}}$, $ M_{0\vert\vec{r}_{1}}$, and $ M_{1\vert\vec{r}_{1}}$, respectively. Based on Alice's measurement, Bob and Charlie select an optimal measurement acting on their own photon to decode a desired bit.

Bob's unsharp measurement can be realized by using
a balanced Mach-Zehnder interferometer (MZI), equipped
with some HWPs and two calcite beam displacers (BDs)
which transmit vertical polarization $\vert V\rangle $ and displace horizontal polarization $\vert H\rangle $ with 3 mm (light green region) \textcolor[rgb]{1,0,0}{\cite{AMC2020,HZH2018,FCT2020,CHP2020,ALY2018}}. In each round of experiment, HWP2 is set at a special angle $\theta_{B} $ to transform the eigenbases, $\lbrace \vert \varphi\rangle, \vert \varphi^{\perp}\rangle \rbrace$, of Bob's current operator to the $\lbrace H, V\rbrace $ basis. Here $ \vert \varphi\rangle $ ($\vert \varphi^{\perp}\rangle$) represents the eigenstate with an outcome equals to 0 (1). More clearly, $ \vert \varphi\rangle\rightarrow \vert H\rangle $ and $\vert \varphi^{\perp} \rangle \rightarrow \vert V\rangle $. Then an unsharp measurement of observable $\Pi= \vert H\rangle\langle H\vert-\vert V\rangle\langle V\vert $ was performed with the elements between HWP2 and HWP4, and finally transformed back to Bob's measurement basis via HWP4 by rotating the same angle as HWP2. A HWP3 spanning across both arms is used to tune the measurement sharpness parameter through changing its angle $\theta_{\lambda} $. For simplicity of the experimental implementation, we only extract the measurement outcome of $ b=0 $ by post-selecting a particular output mode of the interferometer (the blocked mode corresponding to $ b=1 $). Thus after the setup, a Kraus operator $K_{0\vert\vec{s}_{y}}= cos(2 \theta_{\lambda})\vert \varphi\rangle\langle \varphi\vert+sin(2 \theta_{\lambda})\vert \varphi^{\perp}\rangle\langle \varphi^{\perp}\vert $ corresponding to the current unsharp measurement $ M_{0\vert\vec{s}_{y}}$ is realized, where $ M_{0\vert\vec{s}_{y}}=K^{\dagger}_{0\vert\vec{s}_{y}}K_{0\vert\vec{s}_{y}} $ and $\eta_{y}=cos(4 \theta_{\lambda}) $.  Likewise, one just need to replace the rotating operation associated with the wave plates (HWP2, HWP4) by $ \vert \varphi\rangle\rightarrow \vert V\rangle $ and $\vert \varphi^{\perp} \rangle \rightarrow \vert H\rangle $, then the other Kraus operator $K_{1\vert\vec{s}_{y}}= sin(2 \theta_{\lambda})\vert \varphi\rangle\langle \varphi\vert+cos(2 \theta_{\lambda})\vert \varphi^{\perp}\rangle\langle \varphi^{\perp}\vert $ of the other required unsharp measurement $ M_{1\vert\vec{s}_{y}}=K^{\dagger}_{1\vert\vec{s}_{y}}K_{1\vert\vec{s}_{y}} $ can also be obtained. Thus, depending on the orientation of these plates, the interferometer can carry out each of the unsharp operators $\lbrace M_{0\vert\vec{s}_{0}}, M_{1\vert\vec{s}_{0}}, M_{0\vert\vec{s}_{1}}, M_{1\vert\vec{s}_{1}} \rbrace $ required by the protocol.

Charlie's measurements are projective, and therefore, his
setup only consists of a HWP preceded by a PBS (light red region). To reduce the number of components, we replaced the group of three consecutive HWPs with a single HWP (dot green box). The equivalent relationship of the angles of these wave plates is shown in Fig. \ref{setup}. A single-mode fiber finally collects Charlie's photons and brings them to a SPAD, the signal of which is correlated with Alice's signal by a coincidence unit.

In our experiment, we choose twelve different values for Bob's two measurement sharpness parameters $\eta_{0} $ and $\eta_{1}$, separately. Both of them are almost equidistantly distributed in $[0,1]$. Based on the new measurement strategy introduced in the previous section, HWP2, HWP4 and HWP5 choose an appropriate angle from $ \lbrace -33.75^{\circ}, -11.25^{\circ}, 11.25^{\circ}, 33.75^{\circ} \rbrace$ based on the initial bit that Bob and Charlie desire to decode when $ P_{AB}\leq P_{AC} $. Otherwise, we should optimize these angles to maximize the minimum value of $\lbrace P_{AB}, P_{AC} \rbrace$ for a given $\lbrace \eta_{0}, \eta_{1}\rbrace $. Special settings of wave plates in case of $\eta_{0}=0.707 $ and $\eta_{0}=\eta_{1}$ are listed in Appendix A. During the encoding-decoding process, each of Alice, Bob and Charlie has two measurement choices, and for each choice, two trials are needed, corresponding to two different outcomes. Thus, for each fixed $\lbrace \eta_{0},\eta_{1} \rbrace$, we have implemented 32 trials for calculating $ P_{AB} $ and 64 trials for calculating $ P_{AC} $. And for each trial, we record the coincidence counting events for 4 s. The total number of coincident events contributing to a complete measurement is about $ 4\times 10^5 $, sufficient to make the statistical errors small, which is about $ 1\times 10^{-3} $ in our experiment.
 

\section*{IV. Experimental results}

The experimental average success probabilities $P_{AB}$ (red marks) and $P_{AC}$ (blue marks) for several values of $\eta_{0}$ and $\eta_{1}$ are shown in Fig. \ref{range}(a). In general, our experimental results are in good agreement with the theoretical predictions. It clearly shown that double outperform QRAC can be obtained when $\lbrace \eta_{0},\eta_{1}\rbrace$  lies in the right side of the intersection line of the light red plane and the light blue plane (green dashed line). And the useful range for one sharpness parameter increases with the increasing of the other sharpness parameter. The measured and theoretical success probabilities are displayed for fixed values of $\eta_{0}=0.707$ in Fig. \ref{range}(b) and of $\eta_{0}=\eta_{1}$ in Fig. \ref{range}(c). The double QRAC inequality violation range denoted as the light red area, which is enlarged by the inserted box. In particular, even when Bob performs approximately projective measurements, $\eta_{0}=\eta_{1}=0.99$, both Bob and Charlie can beat the classical limit with $P_{AB}=0.77153\pm 0.00078 $ and $P_{AC}=0.77093\pm 0.00071$, which is impossible in the original protocol \cite{MTB2019,FCV2020,AMC2020}. 

In order to illustrate how our measurement strategy works,
we further quantify the degree of measurement sharpness, biasness and incompatibility with the observed two QRAC success probabilities, which can be expressed as:
\begin{equation}\label{sbd1}
\begin{split}
&\eta\geq\eta_{low}=\sqrt{2}(2P_{AB}-1),\\
&\eta\leq \eta_{up}=2\sqrt{(2+\sqrt{2}-4P_{AC})(2P_{AC}-1)},\notag\\
\end{split}
\end{equation}
\begin{equation}\label{sbd2}
\begin{split}
&\vert\vec{s}_{0}\cdot\vec{s}_{1}\vert\leq s_{up}=\dfrac{8P_{AB}-4}{\eta_{1}+\eta_{2}}\cdot\sqrt{2-(\dfrac{8P_{AB}-4}{\eta_{1}+\eta_{2}})^{2}}, \\
&D(\vec{s}_{0}\cdot\vec{s}_{1})\geq 8P_{AB}-6,\\
&D(\vec{t}_{0}\cdot\vec{t}_{1})\geq \dfrac{16P_{AC}-8}{m}-2,\\
\end{split}
\end{equation} 
where $ m $ represents the maximum distance between Charlie's corresponding conditional states. And we can obtain the upper bound of Charlie's measurement biasness $\vert\vec{t}_{0}\cdot\vec{t}_{1}\vert$ by substituting $ \alpha=arccos(s_{up})/2 $ into Eq. (\ref{optiprob}). 
Theoretically, $\vert\vec{s}_{0}\cdot\vec{s}_{1}=cos(2\alpha)$ and $\vert\vec{t}_{0}\cdot\vec{t}_{1}\vert=cos(2\beta)$.  The detailed theoretical calculation is shown in Appendix B. Fig. \ref{range}(d) plots the lower and upper bounds of Bob's sharpness parameter as a function of targeted sharpness $\eta_{t}$. The tightness of the bounds proof that our setup realized a perfect POVM measurement. With the determined sharpness bounds, we can deduce the upper bounds of biasness parameters, $\vert\vec{s}_{0}\cdot \vec{s}_{1}\vert$ and $\vert\vec{t}_{0}\cdot \vec{t}_{1}\vert$, in Bob's and Charlie's respective pair of measurements, which are also shown in Fig. \ref{range}(d). As expected, the measurement biasness parameters become higher than zero near $ \eta = 0.707 $, demonstrating Bob and Charlie need to optimizing their measurement direction when $P_{AB}>P_{AC} $. We also quantify the lower bound degree of incompatibility obtained from the seven experimentally measured probability pairs represented as red lower triangle and blue upper triangle in Fig \ref{range}(c). Fig. \ref{range}(e) presents its relationship with the measurement sharpness and biasness. Obviously, all of the degrees of incompatibility in Bob's and Charlie's respective pair of measurements are above zero verifying measurement incompatibility is necessary to observed outperforming QRAC \cite{TU2020,DSF2020,CHT2020}.


\section*{V. Applications}
\begin{figure}[!h]
\centering\includegraphics[width=9cm]{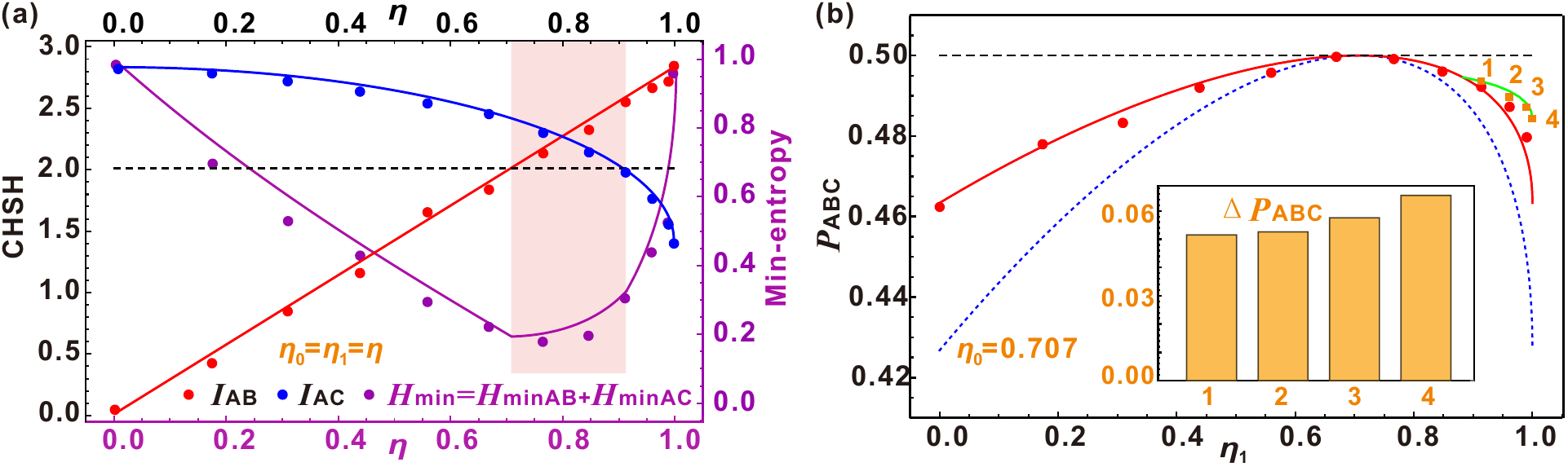}
\caption{(a). Device-independent quantum random number expansion by entanglement-assisted sequential QRAC. The light red region represents the double CHSH inequality violation. (b). The decoding probability of Alice's entire input sequence as a function of $\eta_{1}$. The red (green) curve indicates the theoretical probability that Bob and Charlie have (have not) further optimizing their measurement directions when $P_{AB}>P_{AC}$.  The insert is the increment of decoded probability when Bob and Charlie further optimizing their measurement direction. The blue dashed curve represents the probability of MTB's measurement strategy. Error bars are estimated by the Poissonian statistics of two-photon coincidences which are too small to be visible.}
\label{application}
\end{figure}

In 2010, Pironio \textit{et al.} proposed a random number expansion protocol, where the generated randomness can be certified by violating the Clauser-Horn-Shimony-Holt (CHSH) inequality and quantified by min-entropy \cite{PAM2010}, which reads as:
\begin{equation}\label{RNC}
\begin{split}
&I_{AB}=\dfrac{1}{n}\sum \limits_{x,y}(-1)^{xy}[N(a=b\vert xy)-N(a\neq b\vert xy)], \\
&=(\eta_{0}+\eta_{1})(cos\alpha +cos\beta), \\
&Hmin_{AB}=1-log_{2}(1+\sqrt{1+\dfrac{I_{AB}^2}{4}}),\\
&I_{AC}=\dfrac{1}{n}\sum \limits_{x,z}(-1)^{xz}[N(a=c\vert xy)-N(a\neq c\vert xz)], \\
&=2\cdot M+N\cdot(\sqrt{1-\eta^{2}_{0}}+\sqrt{1-\eta^{2}_{1}}), \\
&Hmin_{AC}=1-log_{2}(1+\sqrt{1+\dfrac{I_{AC}^2}{4}}), 
\end{split}
\end{equation}
where $ M $ and $ N $ have been defined in Eq. (\ref{optiprob}), $N(a = b\vert xy)$ is the number of times that the measurements $\lbrace x, y \rbrace $ are performed with outcomes $a$ and $b$ were found equal after $n$ runs experiment, $N(a \neq b\vert xy)$, $N(a =c\vert xz)$, $N(a \neq c\vert xz)$ are defined analogously. Fig.   \ref{application}(a) confirms the validity of Eq.(\ref{RNC}) in the cases of $\eta_{0}=\eta_{1}=\eta $ and $\alpha=\beta=\pi/4 $. The observed double CHSH inequality violation implies that the positive amount of randomness can be generated not only between Alice and Bob but also between Alice and Charlie at same time. The total min-entropy bound based on entanglement-assisted sequential QRAC as the function of the measurement sharpness is plotted in Fig. \ref{application}(a). The result shows that we can always get the positive amount of randomness as soon as $ \eta >0 $. And the amount of randomness reaches the minimum value during the double CHSH inequality violation range and achieves the maximum number at the full no-signaling condition, i.e., $\eta\rightarrow 0 $ or $\eta\rightarrow 1 $. It means no-signaling condition is critical for guaranteeing the randomness, conversely, the decreasing of the total min-entropy may imply the increasing signaling between Bob and Charlie.


The other application is to increase the decoding probability of Alice's entire input string. When Bob and Charlie cooperate in a communication scenario and agree to always decode different bits, the joint probability of both being correct is given by:
\begin{equation}\label{decode}
\begin{split}
&P_{ABC}=\dfrac{1}{8}\sum \limits_{x,y}P(x_{y}= m_{1}\oplus b,x_{z}= m_{1}\oplus c \vert x,y,z\neq y),\\
&=\dfrac{1}{4}+\dfrac{cos(\alpha + \beta)}{8}(cos\alpha - 
    sin\alpha)+ \dfrac{\eta_{0} + \eta_{1}}{16} (cos\alpha      + sin\alpha) \\
    & + \dfrac{1}{16}(\sqrt{1 - \eta_{0}^2} + \sqrt{
    1 -\eta_{1}^2)} (cos\alpha + 
    sin\alpha) sin(\alpha + \beta)
\end{split}
\end{equation}
Fig. \ref{application}(b) shows the relationship of $P_{ABC}$ versus $\eta_{1}$ when $\eta_{0}=0.707$. It is obvious that our measurement strategy(red and green line) does allow them to obtain a higher success probability than MTB's (blue dashed line). And the increment can further improved by optimizing the decoding direction when $P_{AB}>P_{AC}$ (see the insert in Fig.\ref{application}(b)).


\section*{VI. Conclusions and Discussion}
In this work, we theoretically develop and experimentally demonstrate that it is feasible to sustain double QRAC inequality violation in a much larger sharpness tunable region using only a pair of two-qubit entanglement state by relaxing the equal sharpness and mutual unbiased measurement limits. Larger useful region means stronger noise robustness, which makes our measurement strategy easier to realize quantum correlation sharing in practice. Specifically, we observe a counter-intuitive result that the double QRAC inequality violation can be achieved even both encoders take approximately projective measurements. With the observed success probabilities, we further present a quantitative analysis of the principle of our measurement strategy. Our protocol is based on entangled photon pairs, which is important for tasks that require certified entanglement. Here, we provide a potential application for full device-independent sequential random number expansion. Additionally, we experimentally demonstrate that our optimal measurement strategy can further increase the success probability for Bob and Charlie are cooperate to decode the entirety of Alice's string. Our results provide a more essential way to understand the connection between quantum correlation, quantum measurement and quantum information tasks. 

It would be interesting to extend our protocol to more general sequential QRAC scenarios either involving more measurement settings or higher dimensions. Another  extension is further allowing the encoder to optimize her measurement settings or further permitting the decoders to share more classical information, thus opening up the possibility for future decoders to use adaptive strategies, which may increase the sequence length of outperformed decoders. We will carry out some experiments in these directions in the near future.

\section*{Acknowledgments}

This work was supported by the National Natural Science Foundation Regional Innovation and Development Joint Fund (Grants No. 932021070), the National Natural Science Foundation of China (Grants No.  912122020), the China Postdoctoral Science Foundation (Grant No. 861905020051), the Fundamental Research Funds for the Central Universities (Grant No.841912027, 842041012, 201961009), the Applied Research Project of Postdoctoral Fellows in Qingdao (Grant No. 861905040045), and the Young Talents Project at Ocean University of China (Grant No. 861901013107). The authors thank Xue Bin An and Hong Wei Li for fruitful discussions.


 \appendix
\section*{Appendix A: Measurement settings of Bob and Charlie in different situations}
In this work, we propose a new measurement strategy to expand the useful sharpness range of Bob's measurement to enable both entanglement-assisted sequential QRAC protocol to go over the classical bound by optimizing Bob's and Charlie's measurement setting parameters $\lbrace\alpha,\beta\rbrace$ based on maximizing the minimum value of their average success probabilities $\lbrace P_{AB}$, $P_{AC}\rbrace$. 

As mentioned in the main text, for a fixed pair of sharpness parameters $\lbrace \eta_{0}, \eta_{1} \rbrace $, if $ P_{AB} \leq P_{AC}$, we need to maximize $P_{AB}$ which can be achieved by setting $\alpha=\pi/4 $ and at the meanwhile, the maximum value of $P_{AC}$ can be obtained at $\beta=\pi/4 $. However, if $ P_{AB} > P_{AC}$, we have to adjust their measurement settings to increase the value of  $ P_{AC}$. We start with $\alpha=\pi/4 $ and then  decrease its value gradually, the value of $ P_{AB} $ will decrease, and the value of $  P_{AC} $ will increase at the meanwhile if $\beta $ is updated to maximize $ P_{AC} $. Thus, with the decreasing of $\alpha $, the minimum value of $ \lbrace P_{AB}, P_{AC} \rbrace $ will continuously increased until $P_{AB}=P_{AC}$. Thus the current values of $\alpha$ and $\beta $ which satisfy $ P_{AB}=P_{AC} $ is the optimal measurement setting for Bob and Charlie in the case of $ P_{AB} > P_{AC}$. The optimal pairs of measurement parameters $\lbrace \alpha , \beta \rbrace$ in different situations are shown in  Table. \ref{unequal1}, Table. \ref{unequal2} and Table. \ref{equal}.


\begin{table}[!htb]
  \centering
 \begin{minipage}{1.0\linewidth}
    \renewcommand\arraystretch{1.3}
      \tabcolsep5pt
\begin{tabular}{|c|c|c|c|c|c|c|c|}
\hline
$\theta_{\lambda}$ & $\eta_{1} $ & $\alpha$ & $\beta$ &$\theta_{\lambda}$ & $\eta_{1} $ & $\alpha$ & $\beta$\\
\hline 
$0^{\circ}$ & $1.000$ & $0.00^{\circ}$ & $0.00^{\circ}$ 
&
$8^{\circ}$& $0.848$ &  $6.82^{\circ}$ & $12.55^{\circ}$\\
\hline 
$1^{\circ}$ & $0.998$ & $0.15^{\circ}$ & $2.00^{\circ}
$ 
&$9^{\circ}$ & $0.809$ & $8.74^{\circ}$ & $17.50^{\circ}
$\\
\hline 
$2^{\circ}$ & $0.990$ & $0.42^{\circ}$ & $3.98^{\circ}
$
&$10^{\circ}$ & $0.766$ & $10.97^{\circ}$ & $19.53^{\circ}
$\\
\hline
$3^{\circ}$ & $0.978$ & $0.94^{\circ}$ &  $5.95^{\circ}
$
&$11^{\circ}$ & $0.719$ & $13.58^{\circ}$ & $21.68^{\circ}
$\\
\hline
$4^{\circ}$ & $0.961$ & $1.67^{\circ}$ &  $7.90^{\circ}
$
&$12^{\circ}$ & $0.669$ & $16.64^{\circ}$ & $24.04^{\circ}
$\\
\hline
$5^{\circ}$& $0.940$ &  $2.62^{\circ}$ &  $9.82^{\circ}$
&$13^{\circ}$ & $0.616$ & $20.31^{\circ}$ & $26.74^{\circ}
$\\
\hline
$6^{\circ}$& $0.914$ &  $3.78^{\circ}$ &  $11.72^{\circ}$
&$14^{\circ}$ & $0.559$ & $24.89^{\circ}$ & $30.03^{\circ}
$\\
\hline
$7^{\circ}$& $0.883$ &  $5.18^{\circ}$ & $13.63^{\circ}$&$15^{\circ}$ & $0.500$ & $31.22^{\circ}$ & $34.63^{\circ}
$\\
\hline
\end{tabular}
\caption{The measurement settings when $\eta_{0}=1.0 $.}
\label{unequal1}
 \end{minipage}
  \end{table}


\begin{table}[!htb]
 \centering
  \label{unequal}
    \begin{minipage}{1.0\linewidth}
    \renewcommand\arraystretch{1.3}
      \centering
      \tabcolsep10pt 
\begin{tabular}{|c|c|c|c|}
\hline
$ \theta_{\lambda}$ & $ \eta_{1} $ & $ \alpha $ & $ \beta $\\
\hline 
$1^{\circ}$& $0.998$ & $15.52^{\circ}$ & $24.32^{\circ}$\\
\hline
$2^{\circ}$& $0.990$ & $17.15^{\circ}$ & $26.49^{\circ}$\\
\hline
$3^{\circ}$& $0.978$ & $19.26^{\circ}$ & $28.76^{\circ}$\\
\hline
$4^{\circ}$& $0.961$ & $22.00^{\circ}$ & $31.21^{\circ}$\\
\hline
$5^{\circ}$& $0.940$ & $25.61^{\circ}$ & $33.94^{\circ}$\\
\hline
$6^{\circ}$& $0.914$ & $30.80^{\circ}$ & $37.29^{\circ}$\\
\hline
$7^{\circ}$& $0.883$ & $42.60^{\circ}$ & $43.76^{\circ}$\\
\hline
\end{tabular}
{
 \caption{The measurement settings when $\eta_{0}=0.707$.}
 \label{unequal2}
}
    \end{minipage}
    \end{table}

\begin{table}[!htb]
  \centering
    \begin{minipage}{1.0\linewidth}
    \renewcommand\arraystretch{1.3}
      \tabcolsep10pt
\begin{tabular}{|c|c|c|c|}
\hline
$ \theta_{\lambda} $ & $\eta_{1} $ & $\alpha$ & $\beta $\\
\hline 
$2^{\circ}$ & $0.990$ & $1.12^{\circ}$ & $7.94^{\circ}$\\
\hline
$3^{\circ}$ & $0.978$ & $2.52^{\circ}$ & $11.85^{\circ}$\\
\hline
$4^{\circ}$ & $0.961$ & $4.53^{\circ}$ & $15.72^{\circ}$\\
\hline
$5^{\circ}$ & $0.940$ & $7.23^{\circ}$ & $19.60^{\circ}$\\
\hline
$6^{\circ}$ & $0.914$ & $10.79^{\circ}$& $23.59^{\circ}$\\
\hline
$7^{\circ}$ & $0.883$ & $15.57^{\circ}$& $27.83^{\circ}$\\
\hline
$8^{\circ}$ & $0.848$ & $22.40^{\circ}$& $32.70^{\circ}$\\
\hline
\end{tabular}
{
\caption{The measurement settings when $\eta_{0}=\eta_{1}=\eta.$}
 \label{equal}
}
    \end{minipage}
  \end{table}

\section*{Appendix B: Quantify the degree of measurement biasness,  incompatibility, and sharpness from two QRACs}

In our entanglement-assisted sequential QRAC protocol, a maximum two-qubit entanglement state $(\vert HH \rangle + \vert VV \rangle )/\sqrt{2}$ is sharing among three parties, Alice, Bob and Charlie, where Alice accesses to one qubit, Bob and Charlie access to the other qubit. Then, they perform some measurements on her (his) state to encode or decode a message, which in turn can be expressed as
\begin{equation}\label{measurement}
\begin{split}
&M_{a\vert\vec{r}_{x}}=[I_{A}+(-1)^a ((x \oplus 1) \sigma_{1}+x\sigma_{3})]/2,\\
&M_{b\vert\vec{s}_{y}}=[I_{B}+(-1)^b \eta_{y} (cos(\alpha)\sigma_{1}+ sin(\alpha)\sigma_{3})]/2,\\
&M_{c\vert\vec{t}_{z}}=[I_{C}+(-1)^c (cos(\beta)\sigma_{1}+ sin(\beta)\sigma_{3})]/2,\\
\end{split}
\end{equation}
where $x= x_{0}\oplus x_{1}$, $x_{0}( x_{1})\in\lbrace0,1\rbrace$ is the bit to be encoded, $\eta_{y}\in[0,1]$ is the sharpness parameter and $M_{b\vert\vec{s}_{y}}=K^{\dagger}_{b\vert\vec{s}_{y}}\cdot K_{b\vert\vec{s}_{y}}$, $ K_{b\vert\vec{s}_{y}} $ is the Kraus operator. 

Depending on the choice of measurement and the observed outcome, Alice encodes her two classical bits, $ \lbrace x_{0}, x_{1} \rbrace$, into one of the four possible pure qubit states $\rho_{a\vert\vec{r}_{x}}=[I_{a}+\vec{p}_{a\vert \vec{r}_{x}}\cdot\vec{\sigma}]/2 $ with equal probability $1/4$, whose Bloch vectors can be expressed as
\begin{equation}\label{state}
\begin{split}
p_{0\vert\vec{r}_{0}}=(1,0,0),\
p_{1\vert\vec{r}_{0}}=(-1,0,0),\\
p_{0\vert\vec{r}_{1}}=(0,0,1),\
p_{1\vert\vec{r}_{1}}=(0,0,-1).
\end{split}
\end{equation}
After each round of measurement, Alice sends one classical bit $ m_{1}= x_{0}\oplus a $ to Bob and Charlie to increase their success decoding probabilities. Because of the symmetry of the initial sharing states, Bob's corresponding condition state will be the same as Alice's encoding state. However, due to the influence of Bob's measurements, the state sharing between Alice and Charlie will be decohered to $\rho_{AC}= \dfrac{1}{2}\sum \limits_{y,b}( I_{A}\otimes K_{b\vert\vec{s}_{y}})\rho_{AB}( I_{A}\otimes K^{\dagger}_{b\vert\vec{s}_{y}})$. Thus Charlie's condition state becomes $ \rho_{a\vert\vec{r}_{x}}= \dfrac{1}{2}\sum \limits_{y,b} K_{b\vert\vec{s}_{y}}\rho_{a\vert \vec{r}_{x}}K^{\dagger}_{b\vert\vec{s}_{y}}=\dfrac{1}{2}[I_{c}+\vec{m}_{a\vert \vec{r}_{x}}\cdot\vec{\sigma}] $, where $\vec{m}_{a\vert \vec{r}_{x}}$ is the corresponding Bloch vector, which reads as

\begin{equation}\label{vector}
\begin{split}
& \vec{m}_{0\vert\vec{r}_{0}}=(F\cdot\sin (2 \alpha ), 0, G\cdot\cos (2 \alpha )), \\
&\vec{m}_{1\vert\vec{r}_{0}}=(-F\cdot\sin (2 \alpha ), 0, -G\cdot\cos (2 \alpha )),\\
& \vec{m}_{0\vert\vec{r}_{1}}=(G\cdot\cos (2 \alpha ),0,F\cdot\sin (2 \alpha )),\\
& \vec{m}_{1\vert\vec{r}_{1}}=(-G\cdot\cos (2 \alpha ),0,-F\cdot\sin (2 \alpha )),\\
\end{split}
\end{equation}
where $F= \frac{1}{4}\cdot (\sqrt{1-\eta_{2}^2}-\sqrt{1-\eta_{1}^2})$, $G= \frac{1}{4} (2+\sqrt{1-\eta_{1}^2}+\sqrt{1-\eta_{2}^2}+(2-\sqrt{1-\eta_{1}^2}-\sqrt{1-\eta_{2}^2})$.

Finally, Bob and Charlie perform appropriate measurements respectively on their own states and compare their measurement results with the bit $ m_{1} $ received from Alice to determine their success. The expected average success probability of Bob and Charlie, $P_{AB} $ and $ P_{AC}$, can be expressed as
\begin{equation}\label{pro}
\begin{split}
&P_{AB}=\dfrac{1}{8}\sum \limits_{x,y}P(x_{y}= m_{1}\oplus b\vert x,y), \\
&=\dfrac{1}{2}+\dfrac{1}{16}[(\vec{p}_{0\vert\vec{r}_{0}}-\vec{p}_{1\vert\vec{r}_{0}})\cdot(\eta_{0}\vec{s}_{0}+\eta_{1}\vec{s}_{1})]\\
&+\dfrac{1}{16}[(\vec{p}_{0\vert\vec{r}_{1}}-\vec{p}_{1\vert\vec{r}_{1}})\cdot(\eta_{0}\vec{s}_{0}-\eta_{1}\vec{s}_{1})], \\
& =\dfrac{1}{8}[(4+(\eta_{0}+\eta_{1}))\cdot(cos\alpha+sin\alpha)],\\
&P_{AC}=\dfrac{1}{8}\sum \limits_{x,z}P(x_{z}= m_{1}\oplus c\vert x,z),\\
&=\dfrac{1}{2}+\dfrac{1}{16}[(\vec{m}_{0\vert\vec{r}_{0}}-\vec{m}_{1\vert\vec{r}_{0}})\cdot(\vec{t}_{0}+\vec{t}_{1})]\\
&+\dfrac{1}{16}[(\vec{m}_{0\vert\vec{r}_{1}}-\vec{m}_{1\vert\vec{r}_{1}})\cdot(\vec{t}_{0}-\vec{t}_{1})],\\
&=\dfrac{1}{8}[4+2\cdot M+N\cdot(\sqrt{1-\eta^{2}_{0}}+\sqrt{1-\eta^{2}_{1}})],\\
\end{split}
\end{equation}  
where $ M=cos^{2}\alpha \cdot cos\beta+sin^{2}\alpha \cdot sin\beta $, $N=cos^{2}\alpha \cdot sin\beta+sin^{2}\alpha \cdot cos\beta $, $ \alpha, \beta\in [0,\pi/4] $. We defined the biasness of each decoder's measurement settings by their overlap $\vec{s}_{0}\cdot\vec{s}_{1}$ and $\vec{t}_{0}\cdot\vec{t}_{1}$. Theoretically, $ \vert\vec{s}_{0}\cdot\vec{s}_{1}\vert=cos(2\alpha)$ and $\vert\vec{t}_{0}\cdot\vec{t}_{1}\vert=cos(2\beta)$ for Bob and Charlie, where $\alpha=\pi/4 $ ($\beta=\pi/4$) corresponding to mutually unbiased measurement. 

It is obvious that the success probability depends on the degree of measurement sharpness and biasness. Therefore, by properly adjusting the measurement biasness parameters, the useful range of sharpness parameters to make both QRACs outperform their classical counterparts can be expanded. Now we would like to establish the relationship between the degree of measurement biasness and the average success probability for both Bob and Charlie, thus we can quantify them from these two QRACs. According to Ref. \cite{TKV2018}, for a pair of our fixed qubit observables, the maximum quantum value of QRAC for Alice and Bob satisfies 
\begin{equation}\label{A2}
\begin{split}
&P_{AB} \leq\dfrac{1}{2}+\dfrac{1}{16}( \sqrt{2\mu+2\nu}+\sqrt{2\mu-2\nu}). \\
\end{split}
\end{equation}
where $ \mu=2(\eta_{0}^{2}+\eta_{1}^{2}) $ and  $ \nu=4\eta_{0}\eta_{1}\vec{s}_{0}\cdot\vec{s}_{1} $. We can easily obtain that
\begin{equation}\label{baisB}
\begin{split}
&\vert\vec{s}_{0}\cdot\vec{s}_{1}\vert\leq\dfrac{8P_{AB}-4}{\eta_{1}+\eta_{2}}\cdot\sqrt{2-(\dfrac{8P_{AB}-4}{\eta_{1}+\eta_{2}})^{2}}. \\
\end{split}
\end{equation}  
Combining Eq. (\ref{optiprob}) and Eq. (\ref{baisB}), we can also quantify the bound of $\vert\vec{t}_{0}\cdot\vec{t}_{1}\vert$,  whose value depends on $ \eta_{0} $, $ \eta_{1} $,$ P_{AB} $, $ P_{AC} $ and $ \vert\vec{s}_{0}\cdot\vec{s}_{1}\vert $.

On the other hand, the degree of measurement incompatibility for Bob and Charlie can be expressed as  $ D(\vec{s}_{0}\cdot\vec{s}_{1})=\vert \eta_{0}\vec{s}_{0}+\eta_{1}\vec{s}_{1} \vert+\vert \eta_{0}\vec{s}_{0}-\eta_{1}\vec{s}_{1} \vert $ and $ D(\vec{t}_{0}\cdot\vec{t}_{1})=\vert \vec{t}_{0}+\vec{t}_{1} \vert+\vert \vec{t}_{0}-\vec{t}_{1} \vert$ respectively, which satisfy following condition \cite{AMC2020}  
\begin{equation}\label{incomp}
\begin{split}
&D(\vec{s}_{0}\cdot\vec{s}_{1})\geq 8P_{AB}-6,\\
&D(\vec{t}_{0}\cdot\vec{t}_{1})\geq \dfrac{16P_{AC}-8}{m}-2,\\
\end{split}
\end{equation} 
where $ m $ represents the maximum value of $ \lbrace\vert \vec{m}_{0\vert\vec{r}_{0}}-\vec{m}_{1\vert\vec{r}_{0}} \vert $,  $ \vert \vec{m}_{0\vert\vec{r}_{1}}-\vec{m}_{1\vert\vec{r}_{1}} \vert \rbrace $. 

In order to get a better upper bound or lower bound for above parameters, we should take all kinds of experimental imperfection into consideration. That's to say, all the parameters in Eq. (\ref{baisB}), and  Eq. (\ref{incomp}) should use experimental data. Since the exact measurement sharpness parameter is unknown to us, we also need to estimate its value with the observed success probability. 
It has been demonstrated that its bounds can be determined  as follows \cite{MTB2019}:  
\begin{equation}\label{sharpness}
\begin{split}
 &\eta\geq \eta_{low}=\sqrt{2}(2P_{AB}-1),\\
& \eta\leq \eta_{up}=2\sqrt{(2+\sqrt{2}-4P_{AC})(2P_{AC}-1)}.
\end{split}
\end{equation}
Then, substituting its minimum value into Eq. (\ref{baisB}) and the maximum value into Eq. (\ref{incomp}) respectively, we can obtain the corresponding upper bound and lower bound for the degree of measurement biasness and incompatibility.

\section*{Appendix C: Error estimation}
In our experiment, the statistics of photon counts are assumed to follow a Poisson distribution. We use a subprogram of Poisson Distribution in Wolfram Mathematica 11.2 to simulate the Poisson distribution. The values of each measurement quantity are then calculated from 50 randomly grouped counting sets, in which the error of the quantity is estimated by the standard deviation (SD). To obtain the error of the final value we want (e.g. $ P_{AB} $, $ P_{AC}$ etc), we sum all the errors of the measured quantities included in the equation to calculate the final value. For example, the error of $ P_{AB} $ and $ P_{AC} $ are calculated as

\begin{equation}\label{SD}
\begin{split}
&SD(P_{AB})=\dfrac{1}{8}\sum \limits_{x,y}SD(P(x_{y}= m_{1}\oplus b\vert x,y)), \\
&SD(P_{AC})=\dfrac{1}{8}\sum \limits_{x,z}SD(P(x_{z}= m_{1}\oplus c\vert x,z)).\\
\end{split}
\end{equation}

\section*{Appendix D: Experimental data} 

Tables. \ref{pro4}$ - $\ref{pro8} give details of the experimental data and results from the main text.

\begin{table}[!h]
  \centering
 \begin{minipage}{1.0\linewidth}
    \renewcommand\arraystretch{1.3}
      \tabcolsep5pt
\begin{tabular}{|c|c|c|c|c|c|c|c|}
\hline
$\theta_{\lambda}$ & $\eta_{low} $ & $\eta_{up} $  & $P_{MB0}$ &$P_{MB1}$ &$P_{MC0}$ & $P_{MC1} $ \\
\hline 
$0^{\circ}$ & $0.997$ & $1.000$ & $0.858$ & $0.856$ & $0.677$& $0.674$\\
\hline 
$2^{\circ}$ & $0.964$ & $0.995$ & $ 0.842$ & $0.840$& 
$0.693$ & $0.696$\\
\hline
$3^{\circ}$ & $0.953$ & $0.976$ & $ 0.841$ & $0.833$& 
$0.730$ & $0.700$\\
\hline
$4^{\circ}$ & $0.945$ & $0.968$ & $0.838$ & $0.830$ & 
$ 0.715$ & $0.728$\\
\hline
$5^{\circ}$ & $0.914$ & $0.942$ & $ 0.836$ & $0.810$& 
$0.738$ & $0.734$\\
\hline
$6^{\circ}$& $0.905$ & $0.913$ & $0.820$ & $0.820$ &$0.735$ & $0.762$ \\
\hline
$7^{\circ}$& $0.882$ & $0.888$ & $0.815$ & $0.810$ &$0.761$ & $0.756$ \\
\hline
$8^{\circ}$& $0.826$ & $0.853$ & $0.791$ & $0.792$ &$0.775$ & $0.762$ \\
\hline
$10^{\circ}$ & $0.757$ & $0.756$ & $0.773$ & $0.762$ &$0.781$ & $0.796$ \\
\hline
$12^{\circ}$ & $0.652$ &$ 0.672 $ & $0.732$ & $0.729$ &$0.794$ & $0.821$ \\
\hline
$14^{\circ}$ & $0.587$ &$ 0.597 $ & $0.710$ & $0.705$ &$0.821$ & $0.816$ \\
\hline
$16^{\circ}$ & $0.411$ &$ 0.452 $ & $0.648$ & $0.643$ &$0.840$ & $0.839$ \\
\hline
$18^{\circ}$ & $0.300$ &$ 0.348 $ & $0.621$ & $0.591$ &$0.826$ & $0.826$ \\
\hline
$20^{\circ}$ & $0.151$ &$ 0.221 $& $0.554$ & $0.553$ &$0.848$ & $0.850$ \\
\hline
$22.5^{\circ}$ & $ 0.015$ &$ 0.017 $& $0.481$ & $0.507$ &$0.843$ & $0.857$ \\
\hline
\end{tabular}
{\caption{Details of the results presented in the main text when the measurement directions of Bob and Charlie both are set at $\alpha=\beta=\pi/4 $. The angle of half-wave plates $\theta_{\lambda}$ is used to tune the sharpness of the Bob's effective measurement whose theoretical value equals to $cos (4\theta_{\lambda})$. $\eta_{low} $ and $\eta_{up} $ are the corresponding experimental sharpness lower and upper bound obtained by Eq. (\ref{sharpness}) with the success average probability listed on the right side. $P_{MB0}$($P_{MC0}$) represents the experimental average probability when Bob (Charlie) successfully guessed Alice's first  bit $ x_{0} $ with the corresponding measurement sharpness. Similarly, $P_{MB1}$ ($P_{MC1}$) represents the experimental average probability when Bob (Charlie) successfully guessed Alice's second bit $ x_{1} $. Error bars are estimated by the Poissonian statistics of two-photon coincidences which are about $ 0.001 $ for the measurement sharpness and about $ 0.002 $ for the average success probability.}
\label{pro4}
}
 \end{minipage}
  \end{table}

\begin{table}[!htb]
  \centering
  \label{pro5}
    \begin{minipage}{1.0\linewidth}
    \renewcommand\arraystretch{1.3}
      \tabcolsep9pt
\begin{tabular}{|c|c|c|c|c|c|}
\hline
$ \theta_{\lambda}$& $\eta$ & $ P_{AB} $ & $ P_{AC} $ & $ P_{AB}^{opt} $& $ P_{AC}^{opt} $\\
\hline 
$2^{\circ}$& $0.990$ & $0.841$ & $0.695$ & $0.753$&$0.751$\\
\hline
$3^{\circ}$ &$0.978$& $0.837$ & $0.715$ & $0.753$& $0.750$\\
\hline
$4^{\circ}$ & $0.961$& $0.834$ & $0.721$ & $0.757$&$0.758$\\
\hline
$5^{\circ}$& $0.940$ & $0.823$ & $0.736$ & $0.760$&$0.760$\\
\hline
$6^{\circ}$& $0.914$ & $0.820$ & $0.749$& $0.762$& $0.761$\\
\hline
$7^{\circ}$ & $0.883$ & $0.812$ & $0.758$& $0.770$&$0.767$\\
\hline
$8^{\circ}$ & $0.848$ & $0.792$ & $0.769$& $0.771$&$0.771$\\
\hline
\end{tabular}
{
\caption{The average success probability when  $\eta_{0}=\eta_{1}=\eta$. $ P_{AB} $ and  $ P_{AC} $ are obtained with $\alpha=\beta=\pi/4 $. $ P_{AB}^{opt} $ and  $ P_{AC}^{opt} $ are obtained when Bob and Charlie further optimizing their measurement directions by maximizing the minimum value of $\lbrace P_{AB},P_{AC} \rbrace $, the corresponding values of $\alpha$ and $\beta$ are shown in Table. \ref{equal}. Error bars are estimated by the Poissonian statistics of two-photon coincidences which are about $ 0.002 $.}
\label{pro5}
}
    \end{minipage}
  \end{table}


\begin{table}[!htb]
 \centering
 \label{pro6}
    \begin{minipage}{1.0\linewidth}
    \renewcommand\arraystretch{1.3}
      \centering
      \tabcolsep5pt 
\begin{tabular}{|c|c|c|c|c|c|c|c|}
\hline
$ \theta_{\lambda}$  & $ P_{AB} $ & $ P_{AC} $ & $ P_{ABC} $ & $ P_{AB}^{opt} $& $ P_{AC}^{opt} $& $ P_{ABC}^{opt} $\\
\hline 
$0^{\circ}$& $0.799$ & $0.749$ & $0.753$& $0.748$& $0.752$& $0.484$\\
\hline
$2^{\circ}$& $0.791$ & $0.755$ & $0.430$& $0.765
$& $0.765$& $0.487$\\
\hline
$4^{\circ}$& $0.786$ & $0.761$ & $0.437$& $0.771
$& $0.773$& $0.490$\\
\hline
$6^{\circ}$& $0.781$ & $0.778$ & $0.442$& $0.778
$& $0.780$& $0.494$\\
\hline
\end{tabular}
{
 \caption{The average success probability when $\eta_{0}=0.707$ and $\eta_{1}=cos(4\theta_{\lambda})$.  $ P_{AB} $, $ P_{AC} $ and $ P_{ABC} $ are obtained with $\alpha=\beta=\pi/4 $. $ P_{AB}^{opt} $, $ P_{AC}^{opt} $ and $ P_{ABC}^{opt} $ are obtained when Bob and Charlie further optimizing their measurement directions by maximizing the minimum value of $\lbrace P_{AB},P_{AC} \rbrace $, the     corresponding values of $\alpha$ and $\beta$ are shown in Table. \ref{unequal2}. Error bars are estimated by the Poissonian statistics of two-photon coincidences which are about $ 0.002 $.}
 \label{pro6}
}
    \end{minipage}
    \end{table}



\begin{table}[!htb]
  \centering
  \label{pro7}
 \begin{minipage}{1.0\linewidth}
    \renewcommand\arraystretch{1.3}
      \tabcolsep5pt
\begin{tabular}{|c|c|c|c|c|c|c|c|}
\hline
$\theta_{\lambda}$ & $\eta$  & $\vert\vec{s}_{0}\cdot\vec{s}_{1}\vert$ & $D(\vec{s}_{0}\cdot\vec{s}_{1})$ &$\vert\vec{t}_{0}\cdot\vec{t}_{1}\vert$ &$D(\vec{t}_{0}\cdot\vec{t}_{1})$ \\
\hline 
$2^{\circ}$ & $0.990$ & $0.999$ & $0.019$ & $0.962$ & $0.030$\\
\hline
$3^{\circ}$ & $0.978$ & $0.996$ & $0.040$ & $0.916$ & $0.036$\\
\hline
$4^{\circ}$ & $0.961$ & $0.987$ & $0.068$& $0.853$ & $0.072$\\
\hline
$5^{\circ}$ & $0.940$ & $0.968$ & $0.101$& $0.775$ & $0.106$\\
\hline
$6^{\circ}$ & $0.914$ & $0.930$& $0.137$& $0.680$ & $0.121$\\
\hline
$7^{\circ}$ & $0.883$ &  $0.856$ & $0.175$ & $0.564$ & $0.218$\\
\hline
$8^{\circ}$ & $0.848$ & $0.710$& $0.214$& $0.416$ & $0.292$\\
\hline
\end{tabular}
{
\caption{The experimental data for Fig. \ref{range}(e) in the main text. Error bars are estimated by the Poissonian statistics of two-photon coincidences which are about $ 0.003 $.}
\label{pro7}
}
 \end{minipage}
  \end{table}



\begin{table}[!htb]
  \centering
  \label{pro8}
 \begin{minipage}{1.0\linewidth}
    \renewcommand\arraystretch{1.3}
      \tabcolsep5pt
\begin{tabular}{|c|c|c|c|c|c|c|c|}
\hline
$\theta_{\lambda}$ & $\eta$  & $I_{AB}$ &$I_{AC}$ &$H_{mim}$ \\
\hline 
$0^{\circ}$ & $1.000$ & $2.855$ & $1.405$ & $0.974$ \\
\hline 
$2^{\circ}$ & $0.990$ & $2.727$ & $1.560$ & $0.540$ \\
\hline
$4^{\circ}$ & $0.961$ & $2.675$ & $1.770$ & $0.454$ \\
\hline
$6^{\circ}$ & $0.913$ & $2.560$ & $1.985$ & $0.320$ \\
\hline
$8^{\circ}$ & $0.848$ & $2.332$ & $2.149$ & $0.211$ \\
\hline
$10^{\circ}$ & $0.766$ & $2.141$ & $2.308$ & $0.194$ \\
\hline
$12^{\circ}$ & $0.669$ & $1.844$ & $2.461$ & $0.237$ \\
\hline
$14^{\circ}$ & $0.559$ & $1.660$ & $2.548$ & $0.310$ \\
\hline
$16^{\circ}$ & $0.438$ & $1.164$ & $2.645$ & $0.414$ \\
\hline
$18^{\circ}$ & $0.309$ & $0.851$ & $2.730$ & $0.545$ \\
\hline
$20^{\circ}$ & $0.173$ & $0.427$ & $2.794$ & $0.711$ \\
\hline
$22.5^{\circ}$ & $0.001$ & $0.047$ & $2.830$ & $0.999$ \\
\hline
\end{tabular}
{\caption{The experimental data for Fig. \ref{application}(a) in the main text. Error bars are estimated by the Poissonian statistics of two-photon coincidences which are about $ 0.002 $.}
\label{pro8}
}
 \end{minipage}
  \end{table}



\end{document}